\documentclass[aps,amsmath,amssymb,pra,reprint]{revtex4-1}
\usepackage[english]{babel}
\usepackage{graphicx}
\usepackage{bm}
\usepackage[usenames]{xcolor}
\usepackage{mathrsfs,verbatim}

\usepackage[plainpages=false,pdfpagelabels,colorlinks=true,linkcolor=red,urlcolor=blue,citecolor=blue,pdftitle={Topological phase transitions in finite-size periodically driven translationally invariant systems},pdfauthor={},pdfdisplaydoctitle=true,pdfduplex=DuplexFlipLongEdge]{hyperref}

\newcommand{\assign}{:=}
\newcommand{\mathd}{\mathrm{d}}
\newcommand{\mathe}{\mathrm{e}}
\newcommand{\mathi}{\mathrm{i}}
\newcommand{\mathpi}{\pi}
\newcommand{\nocomma}{}
\newcommand{\nosymbol}{}

\newcommand{\tmacronym}[1]{\textsc{#1}}
\newcommand{\tmem}[1]{{\em #1\/}}
\newcommand{\tmmathbf}[1]{\ensuremath{\boldsymbol{#1}}}
\newcommand{\tmop}[1]{\ensuremath{\operatorname{#1}}}

\newcommand{\ygllangle}{\ensuremath{\langle\hskip-0.2em\langle}}
\newcommand{\ygrrangle}{\ensuremath{\rangle\hskip-0.2em\rangle}}

\begin{document}

\title{Topological phase transitions in finite-size periodically driven\\ translationally invariant systems}

\author{Yang Ge}
\author{Marcos Rigol}
\affiliation{Department of Physics, The Pennsylvania State University, University Park, PA 16802, USA}

\begin{abstract}
It is known that, in the thermodynamic limit, the Chern number of a translationally invariant system cannot change under unitary time evolutions that are smooth in momentum space. Yet a real-space counterpart of the Chern number, the Bott index, has been shown to change in periodically driven systems with open boundary conditions. Here we prove that the Bott index and the Chern number are identical in translationally invariant systems in the thermodynamic limit. Using the Bott index, we show that, in finite-size translationally invariant systems, a Fermi sea under a periodic drive that is turned on slowly can acquire a different topology from that of the initial state. This can happen provided that the gap-closing points in the thermodynamic limit are absent in the discrete Brillouin zone of the finite system. Hence, in such systems, a periodic drive can be used to dynamically prepare topologically nontrivial states starting from topologically trivial ones.
\end{abstract}

\maketitle

\section{Introduction}

Topological insulators have attracted much attention in the last decade \cite{Hasan2010, Qi2011}. While they might appear as equilibrium phases in some materials, applying a time varying potential provides a flexible way to induce topological phases in insulators that are topologically trivial otherwise. In particular, a system driven periodically in time can exhibit so-called Floquet topological phases \cite{Oka2009, Lindner2011, Kitagawa2010}. For example, a high-frequency periodic drive can modify the topological structure of energy bands giving rise to a rich realm of exotic states \cite{Kitagawa2011, Mikami2016, Du2017}. A few experiments have been carried out to explore topological phases in periodically driven systems \cite{Rechtsman2013,Jotzu2014,Huang2017}. Closely related to the model studied here, in Ref.~\cite{Jotzu2014} a driven Fermi sea of ultracold atoms in a honeycomb lattice acquired a nontrivial topology as predicted by the Haldane model \cite{Haldane1988}.

The unitary time evolution of the topological properties of a Fermi sea, as it turns out, is fundamentally different from just tuning the Floquet Hamiltonian across different phases. In Ref.~\cite{DAlessio2015}, a no-go theorem was proved for dynamics under a simple two-band Hamiltonian in two dimensions. It states that the topological index, the Chern number, of a Fermi sea that is in a pure state does not change during unitary dynamics under rather general conditions. The system studied consisted of spinless fermions on an infinite translationally invariant (periodic boundary conditions) honeycomb lattice. As a result, the Hamiltonian is block diagonal in the crystal momentum space. At each crystal momentum $\mathbf{k}$, the Bloch Hamiltonian has the form $\hat{H}_\mathbf{k} (t) = - \vec{B}_\mathbf{k} (t) \cdot \vec{\sigma}$, where $\vec{\sigma}$ are the Pauli matrices and $\vec{B}_\mathbf{k} (t)$ is the pseudomagnetic field. If the initial state is pure and both the pseudomagnetic field and the pseudospin are smooth in $\mathbf{k}$ space, then by the no-go theorem the first Chern number is a constant of motion \cite{DAlessio2015, Caio2015}, as seen in studies of quantum quenches \cite{Foster2013, Foster14, Sacramento2014}. This implies that, in infinite systems, an adiabatic annealing that changes the Chern number can never be achieved \cite{Privitera2016}.

Much work has been done to explain the change in topology as observed in experiments. One approach is to embed the out-of-equilibrium system in a dissipative setting. This can be achieved by introducing a thermal bath \cite{Dehghani2015, Dehghani2014} or dephasing noise \cite{Hu2016}. Non-unitary evolutions destroy coherence within the quantum state. Consequently, the Hall conductance and a generalized version of the Chern number can change \cite{Bardyn2013, Hu2016}. Similar results can be obtained in the context of the diagonal ensemble \cite{Wang2016a, Wang2016b}. Another important point is that equivalent formulations of a topological index in equilibrium may not be equivalent any more out of equilibrium, and the response function of the system can be contained in a non-conserved formulation \cite{Foster2013, Wang2016b, Caio2016, Wilson2016}.

Intriguingly, boundary conditions also appear to determine whether a topological index can change under a periodic drive, as demonstrated in the study of systems with open boundary conditions~\cite{DAlessio2015}. In those systems, which lack translational symmetry, the Bott index is a topological index that can be used in place of the Chern number. The Bott index is defined in real space and does not require transitional invariance \cite{Loring2010, Hastings2011, Titum2015}. In Ref.~\cite{DAlessio2015}, when evolving a finite Fermi sea under a periodic drive that was turned on slowly, it was found that the Bott index can change from a value determined by the initial Hamiltonian to that of the equilibrium Floquet bands. Thus, for open boundary conditions (the case in experiments), the Bott index is not a conserved quantity.

The contrast between the no-go theorem for infinite translationally invariant systems and the fact that the Bott index can change under unitary dynamics in finite lattices with open boundary conditions motivates us to further explore the relation between the Chern number and the Bott index and to study the time evolution of the Bott index in finite lattices with periodic boundary conditions. First, we reformulate the Bott index in momentum space and prove that it is equivalent to the Chern number in the thermodynamic limit. When written in momentum space, the Bott index is nothing but the integer formulation of the Chern number in finite lattices as derived in Ref.~\cite{Fukui2005} from lattice gauge theory. In addition to being a gauge-independent integer by definition, this topological index has the advantage that, with increasing system size, it converges to the thermodynamic result much more rapidly than the usually used discretized integration of the traditional Chern number. Our second goal is to understand the dynamics of the Bott index in finite translationally invariant systems under the same model Hamiltonian as in Ref.~\cite{DAlessio2015}. We show that the Bott index can change in incommensurate lattices, as it does in systems with open boundary conditions. There is a finite time scale for the turn on of the periodic drive that enables this topological transition to occur. This time scale diverges with increasing system size, as expected from the no-go theorem for infinite systems.

The presentation is organized as follows. In Sec.~\ref{sec-bott-eq-ch}, we reformulate the Bott index in momentum space and prove its equivalence with the Chern number in the thermodynamic limit. In Sec.~\ref{ss-steady}, we introduce the model Hamiltonian, and use the Bott index and a finite-size scaling analysis in translationally invariant lattices to determine the phase diagram of the Floquet Hamiltonian in the thermodynamic limit. The dynamical behavior of the Bott index in finite translationally invariant systems is studied in Sec.~\ref{sec:dynbott}. A summary of our results is presented in Sec.~\ref{sec:summary}.

\section{Equivalence between the Bott index and the Chern number}\label{sec-bott-eq-ch}

The Chern number of an energy band is defined as the integral of the Berry curvature over the Brillouin zone (BZ) \cite{Hasan2010,Thouless1982}
\begin{equation}\label{Ch}
  \tmop{Ch} (n) = \frac{\mathi}{2 \mathpi} \int_{\tmacronym{\tmop{BZ}}} \mathd^2 k \ \nabla \times \tmmathbf{\mathcal{A}}_n \left( \mathbf{k} \right),
\end{equation}
where $n$ is the band index, $\tmmathbf{\mathcal{A}}_n \left( \mathbf{k} \right) \assign \langle u_n \left( \mathbf{k} \right) | \nabla_{\mathbf{k}} \left| u_n \left( \mathbf{k} \right) \right\rangle$ is the Berry connection, and $\left| u_n \left( \mathbf{k} \right) \right\rangle$ is the eigenstate of the Bloch Hamiltonian at crystal momentum $\mathbf{k}$ in the $n$th band. By definition, the Chern number is a topological index for translationally invariant two-dimensional (2D) systems. For 2D systems that lack translational invariance, one can use the Bott index introduced by Loring and Hastings \cite{Loring2010} as the analog of the Chern number. This was done in Ref.~\cite{DAlessio2015} to study topological properties of Fermi seas in patch geometries and their unitary time evolution under a periodic drive.

Consider a 2D lattice spanned by primitive (right-handed) vectors $\vec{a}_\mu, \mu = 1$ and $2$, with $L_\mu$ lattice sites along each $\vec{a}_\mu$ such that it has $L_1 \times L_2$ lattice sites. Let $l_\mu \in [0,L_\mu)$ be the spatial coordinate along $\vec{a}_\mu$ such that a lattice site $(l_1, l_2)$ is at $l_1 \vec{a}_1 + l_2 \vec{a}_2$. One can define the operators $\hat{U} \assign \mathe^{2 \mathpi \mathi \hat{l}_1 / L_1}$ and $\hat{V} \assign \mathe^{2 \mathpi \mathi \hat{l}_2 / L_2}$. To study a Slater determinant state its projection operator $\hat{P}$ is used to produce the reduced matrix $\tilde{U}$ given by $\hat{P} \hat{U} \hat{P} \circeq \left( \begin{smallmatrix} 0&0\\0&\tilde{U} \end{smallmatrix} \right)$ in the basis of occupied (lower right) and unoccupied (upper left) single particle states, and similarly $\tilde{V}$. The Bott index for such a state is given by
\begin{equation}\label{bott-def}
  C_b (\hat{P}) = \frac{1}{2 \mathpi} \tmop{Im} \tmop{Tr} \ln (\tilde{V} \tilde{U} \tilde{V}^{\dag}  \tilde{U}^{\dag}).
\end{equation}
The Bott index is well defined in systems for which the Chern number is not. It requires neither translational invariance nor completely filled energy bands. Yet the Bott index of a conducting state is ill defined because the corresponding $\tilde{V} \tilde{U} \tilde{V}^{\dag} \tilde{U}^{\dag}$ matrix is singular \cite{Loring2010}. As with the Chern number, a topologically trivial state has zero Bott index, and a nonzero Bott index counts the number of topologically protected edge modes at the boundaries of the system \cite{Titum2015}.

In finite translationally invariant systems, one can rewrite the Bott index in crystal momentum space. Here we use the coordinate system $(k_1, k_2)$, with $k_\mu \in [0,2\pi/a_\mu)$ being the component along primitive reciprocal vectors $\vec{b}_\mu$, for which $\vec{a}_\mu \cdot \vec{b}_\nu = 2\pi \delta_{\mu \nu}$. Our set of $\mathbf{k}$ points of interest is within the parallelogram bounded by $\vec{b}_1$ and $\vec{b}_2$, which is equivalent to the first Brillouin zone. The infinitesimal momentum space distances between neighboring momentum space points are $\mathord{\delta \mathbf{k}}_1 = [2 \pi / (L_1 a_1), 0]$ and $\mathord{\delta \mathbf{k}}_2 = [0, 2 \pi /( L_2 a_2)]$. Let $\mathbf{q}_0 \assign \mathbf{k}$, $\mathbf{q}_1 \assign \mathbf{k} - \mathord{\delta \mathbf{k}}_1$, $\mathbf{q}_2 \assign \mathbf{k} - \mathord{\delta \mathbf{k}}_2$, $\mathbf{q}_3 \assign \mathbf{k} - \mathord{\delta \mathbf{k}}_1 - \mathord{\delta \mathbf{k}}_2$. The operators $\hat U$ and $\hat V$ are infinitesimal translation operators in momentum space, i.e., $\langle \psi_n \left( \mathbf{k} \right) | \hat U \left| \psi_m \left( \mathbf{k}' \right) \right\rangle = \left\langle u_n \left( \mathbf{k} \right) \middle| u_m \left( \mathbf{k}' \right) \right\rangle \delta_{\mathbf{q}_1, \mathbf{k}'}$ and $\langle \psi_n \left( \mathbf{k} \right) | \hat V \left| \psi_m \left( \mathbf{k}' \right) \right\rangle = \left\langle u_n \left( \mathbf{k} \right) \middle| u_m \left( \mathbf{k}' \right) \right\rangle \delta_{\mathbf{q}_2, \mathbf{k}'}$, for a Bloch state $\left| \psi_n \left( \mathbf{k} \right) \right\rangle$ normalized as $\left\langle \psi_n \left( \mathbf{k} \right) \middle| \psi_{n'} \left( \mathbf{k}' \right) \right\rangle = \delta_{n \nocomma n'} \delta_{\mathbf{k} \nocomma \mathbf{k}'}$. 

In the Bloch-state basis, the matrix elements of $\tilde{V} \tilde{U} \tilde{V}^{\dag} \tilde{U}^{\dag}$ can be written as
\begin{equation}\label{vuvu-element}
  \langle \psi_n \left( \mathbf{k} \right) | \tilde{V} \tilde{U} \tilde{V}^{\dag}  \tilde{U}^{\dag} \left| \psi_{n'} \left( \mathbf{k}' \right) \right\rangle = \delta_{\mathbf{k} \nocomma \mathbf{k}'} \sum_{j \nocomma l \nocomma m} \mathcal{U}_{n \nocomma j}^{0 \nocomma 2} \mathcal{U}_{j \nocomma l}^{2 \nocomma 3} \mathcal{U}_{l \nocomma m}^{3 \nocomma 1} \mathcal{U}_{m \nocomma n'}^{1 \nocomma 0},
\end{equation}
where $\mathcal{U}_{n \nocomma j}^{\alpha \nocomma \beta} \assign \left\langle u_n \left( \mathbf{q}_\alpha \right) \middle| u_j \left( \mathbf{q}_\beta \right) \right\rangle$, with $\alpha,\beta=0$, 1, 2, and 3. Thus, $\tilde{V} \tilde{U} \tilde{V}^{\dag} \tilde{U}^{\dag}$ is block diagonal in momentum space. The indices $j, l, m$ run over filled bands for a given $\mathbf{k}$.

For a single band, we then have that
\begin{equation}\label{bott-single-body}
  {C_b} (n) = \frac{1}{2 \mathpi}  \sum_{\mathbf{k} \in \tmacronym{\tmop{BZ}}} \tmop{Im} \ln (\mathcal{U}_{n \nocomma n}^{0 \nocomma 2} \mathcal{U}_{n \nocomma n}^{2 \nocomma 3} \mathcal{U}_{n \nocomma n}^{3 \nocomma 1} \mathcal{U}_{n \nocomma n}^{1 \nocomma 0})
\end{equation}
This expression was derived for the Chern number in finite translationally invariant systems in Ref.~\cite{Fukui2005}. As discussed there, the result of Eq.~\eqref{bott-single-body} in finite systems converges much faster to the value of the Chern number in the thermodynamic limit than the discretized version of Eq.~\eqref{Ch}. In addition, ${C_b} (n)$ is gauge independent.

The Bott index was shown to give the Hall conductance in Ref.~\cite{Hastings2011}. Hence, it is equivalent to the Chern number. Below, we give an elementary proof that, in the thermodynamic limit, the Bott index is identical to the Chern number. The only requirement for this proof is that the occupied single-particle Bloch states be locally $\mathcal{\mathbb{C}}^{2}$ in momentum space.

First, we expand $\left| u_n \left( \mathbf{q}_\alpha \right) \right\rangle$, with $\alpha=1$, 2, and 3, about $\mathbf{k}$. It gives 
\begin{equation}
 \left| u_n \left( \mathbf{q}_1 \right) \right\rangle = \left| u_n \right\rangle - \mathord{\delta k}_1 \frac{\partial | u_n \rangle}{\partial k_1} + \frac{\left( \mathord{\delta k}_1 \right)^2}{2}  \frac{\partial^2 | u_n \rangle}{\partial k_1^2} + O \left( \mathord{\delta k}_1^3 \right),
\end{equation}
where, on the right-hand side, we omitted the momentum argument as all kets and their derivatives are evaluated at $\mathbf{k}$ (we follow this convention in the expressions below). Similarly, one can expand $\left| u_n \left( \mathbf{q}_2 \right) \right\rangle$ and $\left| u_n \left( \mathbf{q}_3 \right) \right\rangle$. Plugging those expansions into Eq.~\eqref{vuvu-element}, one finds that the $n\neq n'$ matrix elements scale as $\langle \psi_n \left( \mathbf{k} \right) | \tilde{V} \tilde{U} \tilde{V}^{\dag} \tilde{U}^{\dag} \left| \psi_{n'} \left( \mathbf{k} \right) \right\rangle \sim O \left[( \mathord{\delta k})^2 \right]$, where we assumed that $\mathord{\delta k}\sim\mathord{\delta k_1}\sim\mathord{\delta k_2}$. For the diagonal entries, on the other hand,
\begin{widetext}
\begin{eqnarray} \label{vuvu-dbar}
  \langle \psi_n \left( \mathbf{k} \right) | \tilde{V} \tilde{U} \tilde{V}^{\dag} \tilde{U}^{\dag} \left| \psi_n \left( \mathbf{k} \right) \right\rangle & = & \mathbf{1} + \delta k_1 \delta k_2  \left( \frac{\partial \langle n|}{\partial k_2}  \frac{\partial | n \rangle}{\partial k_1} - \mathrm{c.c.} \right) + \delta k_1 \delta k_2  \sum_m \left[ \langle n|  \frac{\partial | m \rangle}{\partial k_1}  \frac{\partial \langle m|}{\partial k_2}  | n \rangle - \mathrm{c.c.} \right]\nonumber\\* &  & + \left[ (\delta k_1)^2  \left( \sum_m \left| \langle n| \frac{\partial | m \rangle}{\partial k_1} \right|^2 - \left| \frac{\partial | n \rangle}{\partial k_1} \right|^2 \right) + (x \rightarrow y) \right] + O (\delta k^3). 
\end{eqnarray}
\end{widetext}

Given those results for the diagonal and off-diagonal matrix elements of $\tilde{V} \tilde{U} \tilde{V}^{\dag} \tilde{U}^{\dag}$, one can evaluate Eq.~\eqref{bott-def} using the fact that, for a general matrix $A$ decomposed as $A = \mathbf{1} + A_{\bar{D}} + A_O$, where $A_{\bar{D}}$ is the diagonal part of $A - \mathbf{1}$ and $A_O$ is the off-diagonal part, one can write
\begin{equation}\label{tr-log-expand}
  \tmop{Tr} \ln A = \tmop{Tr} A_{\bar{D}} + \tmop{Tr} [O (A^2_{\bar{D}}, A^2_O, A_{\bar{D}} A_O)].
\end{equation}
Since $(\tilde{V} \tilde{U} \tilde{V}^{\dag} \tilde{U}^{\dag})_{\bar{D}}$ and $(\tilde{V} \tilde{U} \tilde{V}^{\dag} \tilde{U}^{\dag})_O$ are order $\delta k^2$ or higher, it follows from Eq.~\eqref{tr-log-expand} that $\tmop{Tr} \ln \tilde{V} \tilde{U} \tilde{V}^{\dag} \tilde{U}^{\dag} = \tmop{Tr} (\tilde{V}  \tilde{U} \tilde{V}^{\dag}  \tilde{U}^{\dag})_{\bar{D}} + \tmop{Tr} \left[ O \left( \mathord{\delta k}^4 \right) \right]$. Taking the imaginary part of the trace gives
\begin{eqnarray}
  \tmop{Im} \tmop{Tr} \ln \tilde{V}  \tilde{U}  \tilde{V}^{\dag} \tilde{U}^{\dag} & = & \frac{1}{\mathi}  \sum_n \sum_{k_1 \nocomma k_2} \mathord{\delta k}_1  \mathord{\delta k}_2  \left( \frac{\partial \langle n|}{\partial k_2} \frac{\partial | n \rangle}{\partial k_1} - \mathrm{c.c.} \right)\nonumber\\* &  & + O \left( \mathord{\delta k} \right). 
\end{eqnarray}
In the limit $L_1, L_2 \rightarrow \infty$, $\sum_{k_1 \nocomma k_2} \mathord{\delta k}_1  \mathord{\delta k}_2 \rightarrow \int_{\tmacronym{\tmop{BZ}}} \mathd^2 k$, and all higher-order terms vanish. The trace becomes the integral of Berry curvature. Therefore, in the thermodynamic limit, for a Fermi sea occupying all $n \leqslant N$ bands
\begin{equation}
  {C_b} (\hat{P}_{n \leqslant N})  =  \sum_{n \leqslant N} {C_b} (n)  =  \sum_{n \leqslant N} \tmop{Ch} (n) . 
\end{equation}
Hence, for a Fermi sea that is locally $\mathbb{C}^2$ in $\mathbf{k}$ space, the Chern number and the Bott index are identical in the thermodynamic limit. Each term of the sum in Eq.~\eqref{bott-single-body} is simply the local Berry curvature times the area element $\delta k^2$, in other words the Berry connection around the boundary of that area element, as shown in Ref.~\cite{wu_gauge-fixed_2012}. If the Fermi level is in the middle of a band, which corresponds to a conducting state, some $\mathbf{k}$ points in the Brillouin zone will have underfilled neighbors $\mathbf{q}_1$ and/or $\mathbf{q}_2$. That $\mathbf{k}$ block is then singular and so is $\tilde{V} \tilde{U} \tilde{V}^{\dag} \tilde{U}^{\dag}$. Thus, in this case the Bott index is ill defined. For Fermi seas with a well defined Bott index or Chern number, these two topological indices are well-defined and equivalent during unitary time evolutions under Hamiltonians that are $\mathbb{C}^2$ in $\mathbf{k}$ space.

\section{Model Hamiltonian and Floquet topological phases}\label{ss-steady}

Having established the equivalence between the Bott index and the Chern number, in what follows we study the dynamics of the Bott index in systems with periodic boundary conditions. Our goal is to understand how it compares to the dynamics of the same topological index in systems with open boundary conditions~\cite{DAlessio2015}.

We consider a tight-binding model of spinless fermions on a honeycomb lattice with nearest-neighbor ($\langle j, l \rangle$) hopping and a sublattice staggered potential at half filling. An in-plane circularly polarized electric field, which is uniform in space, provides the time-periodic drive. In units of $\hbar = 1$, the Hamiltonian is
\begin{equation} \label{hamiltonian}
  \hat H (t)  =  - J \sum_{\langle j, l \rangle} \left[\mathe^{\mathi e \vec{A} (t)  \vec{d}_{j \nocomma l}} \hat c_j^{\dag} \hat c_l + \text{H.c.} \right]+ \frac{\Delta}{2}  \sum_{\substack{j \in \mathcal{A}\\ l \in \mathcal{B}}} (\hat n_j -_{\nosymbol} \hat n_l ). 
\end{equation}
The 2D vector potential $\vec{A} (t) = A~(\sin \Omega t, \cos \Omega t)$ accounts for the electric field. It introduces a phase when particles hop from site $l$ to one of its nearest-neighbors sites $j$, separated by a distance $d=|\vec{d}_{jl}|$. The second term in $\hat H (t)$, with site number operators $\hat n_j$, describes the staggered potential (of strength $\Delta$) between the $\mathcal{A}$ and $\mathcal{B}$ sublattices in the honeycomb lattice. In a translationally invariant system, this Hamiltonian is block diagonal in momentum space. Each momentum block is described by a pseudomagnetic field $-\vec{B}_\mathbf{k} \cdot \vec{\sigma}$ acting on the sublattice spinor $(\hat{c}_{\mathbf{k}, \mathcal{A}}, \hat{c}_{\mathbf{k}, \mathcal{B}})^{\small{T}}$.

When both $A$ and $\Delta$ are zero, the energy bands are gapless at $K$ and $K'$ in the Brillouin zone. For convenience, we set the lattice constants $a_\mu$ to 1, and the coordinates of $K$ and $K'$ are $(\frac{2 \pi}{3},\frac{4 \pi}{3})$ and $(\frac{4\pi}{3},\frac{2\pi}{3})$, respectively. Those band-touching points are protected by the combination of inversion symmetry and time-reversal symmetry. In the static case ($A=0$), a nonzero $\Delta$ introduces a $B_z$ of equal magnitude at $K$ and $K'$, still related by time-reversal symmetry, and opens a gap. In this work, we set $\Delta = 0.15 J$ in order to be close to the experimental parameters in Ref.~\cite{Jotzu2014}. Both static bands have zero Chern numbers; that is, they are topologically trivial.

The time-dependent electric field breaks time-reversal symmetry. Its effect is manifest in the Floquet picture, which follows after the Floquet theorem. The Floquet theorem states that, for a Hamiltonian that is periodic in time $\hat H (t + T) = \hat H (t)$, the evolution operator over one period can be written as
\begin{equation}
  \hat U (t_0 + T, t_0) = \exp [- \mathi \hat H_F (t_0) T],
\end{equation}
where $\hat{H}_F (t_0)$ is the time-independent Floquet Hamiltonian (it depends, in general, on the selected starting time $t_0$ defining the period) \cite{Shirley1965,Sambe1973,Bukov2015}. The eigenstates of $\hat H_F$ are stationary states of the driven system at stroboscopic times.

Under high driving frequencies $\Omega=2\pi/T$, $\hat H_F$ can be extracted from a high-frequency expansion \cite{Mikami2016, Bukov2015}. To $O (\Omega^{- 1})$, the rotating electric field renormalizes the nearest-neighbor hopping amplitude and induces next-nearest-neighbor ($\ygllangle j, l \ygrrangle$) hoppings. The Floquet Hamiltonian reads
\begin{eqnarray}
  \hat H_{F} (t)  & = & - J \mathcal{J}_0(e a A) \sum_{\langle j, l \rangle} \left(\hat c_j^{\dag} \hat c_l + \text{H.c.} \right) \nonumber \\ && + \frac{J^2}{\Omega} \sum_{\ygllangle j, l \ygrrangle} \left(\mathi K_{jl} \hat c_j^{\dag} \hat c_l + \text{H.c.} \right) \nonumber\\* &   & + \frac{\Delta}{2}  \sum_{\substack{j \in \mathcal{A}\\ l \in \mathcal{B}}} (\hat n_j -_{\nosymbol} \hat n_l ) + O ( \Omega^{-2}),
\end{eqnarray}
where $\mathcal{J}_n$ are the Bessel functions of the first kind, and
\begin{equation}
K_{jl}=s_{jl} \sum_{n=1}^\infty \frac{2}{n} \mathcal{J}_n^2 ( e a A ) \sin \frac{2 n \pi}{3}.
\end{equation}
The sign $s_{jl}$ is $+$ ($-$) if the two-step hopping, going around the hexagon corners, has the same (opposite) chirality as the polarization of the electric field.

\begin{figure}[!t]
  \includegraphics[width=0.99\columnwidth]{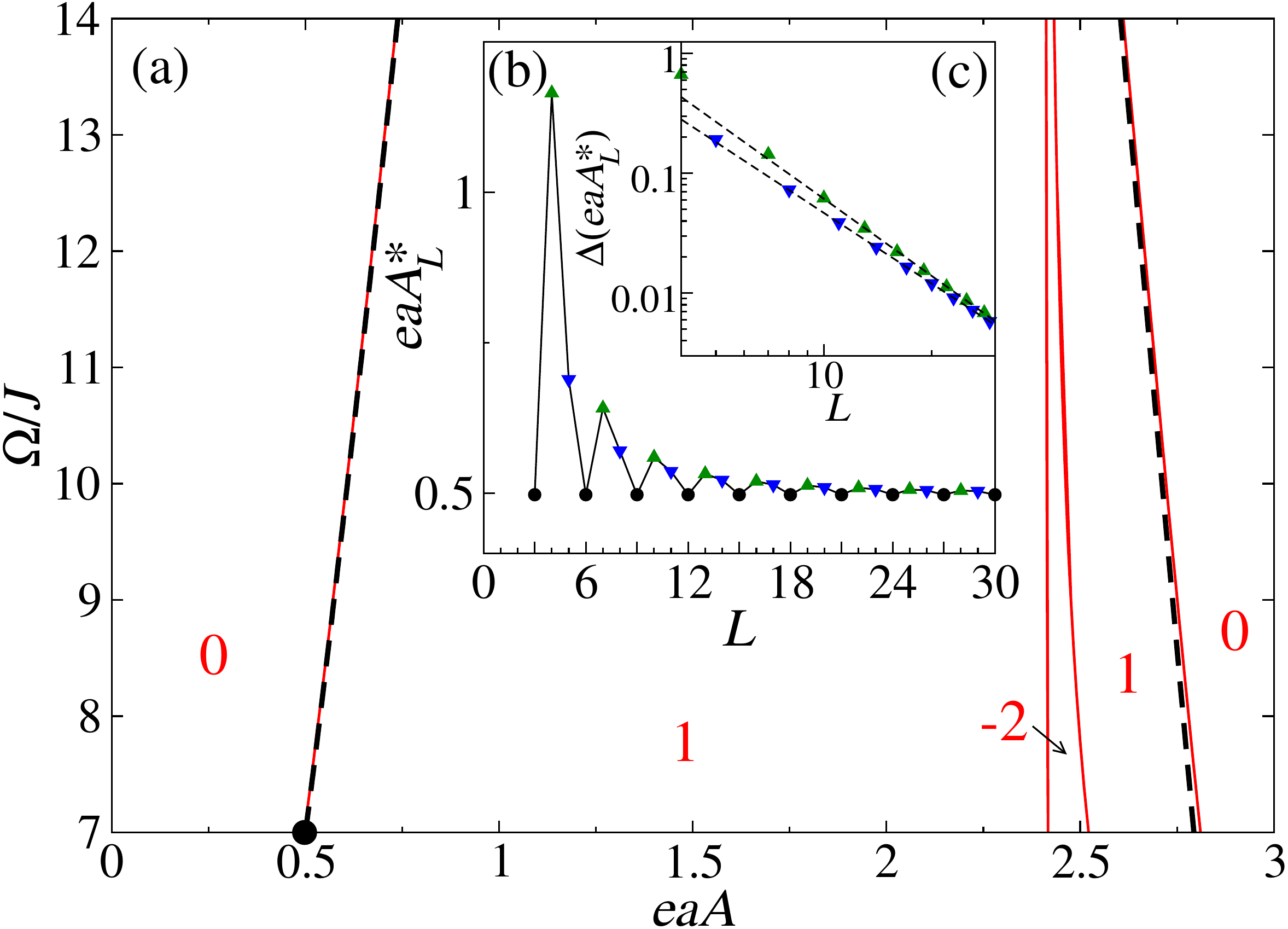}
  \caption{(a) Chern number phase diagram in the driving frequency $\Omega$ and $eaA$ plane for $\hat H_F$ obtained from a high-frequency expansion to $O(\Omega^{-1})$ (dashed lines) and from numerically exact calculations (solid lines). The Chern number is computed using the Bott-index formula in Eq.~(\ref{bott-def}) in finite commensurate systems that are sufficiently large so that the result does not change (within machine precision) with increasing system size. (b) Critical value of the magnitude of the electric field $eaA_L^{\ast}$ for the first topological transition when $\Omega = 7 J$ [black dot in (a)] in systems with $L_1 \times L_2$ lattice sites plotted as a function of $L_1=L_2\equiv L$. (c) Scaling of the critical value for incommensurate systems. We plot $\Delta (eaA_L^{\ast}) \assign ea (A_L^{\ast} - A^{\ast}_{\infty})$ vs $L$ for lattices with $L=3 \iota + 1$ and $L=3 \iota + 2$ ($\iota\in \mathcal{\mathbb{Z}}$), as well as fits to $ \Delta (eaA_L^{\ast}) = \gamma\, L^{-2 x}$ for $L \ge 10$. The fits yield $x \approx 1.07$ and $0.98$, respectively.}\label{fig-static}
\end{figure}

In momentum space, next-nearest neighbor hoppings contribute to $B_z$ of the Hamiltonian at $K$ and $K'$. For $0 < e a A < 1.69$, its sign is the opposite to (same as) $B_z$ generated by $\Delta$ at $K'$ ($K$). As a result, the Floquet band gap closes at $K'$ upon increasing the magnitude of the vector potential. This results in a topological phase transition in which the Chern number changes from 0 to 1 [see dashed line in Fig.~\ref{fig-static}(a)]. A second topological phase transition in which the Chern number changes from 1 to 0, and the gap closes once again at $K'$ occurs upon further increasing the magnitude of the vector potential.

Higher order terms in $\Omega^{- 1}$ introduce further neighbor hoppings that can, in turn, generate new topological phase transitions if $\Omega$ is not too large. In Fig.~\ref{fig-static}(a), we show the Chern number phase diagram (solid lines) obtained using a numerically exact calculation of $\hat U (T,0)$ \cite{DAlessio2015}. One can see that, as a result of terms $O(\Omega^{- 2})$ and higher, two additional phase transitions appear in the regime studied. Interestingly, the corrections to the critical values obtained for the transitions between phases with Chern numbers of 0 and 1 are small. A detailed discussion of the phase diagram of the system studied here, for $\Delta=0$, can be found in Ref.~\cite{Mikami2016}. We note that the Chern number of a Floquet system does not directly give the number of topologically protected edge states \cite{Rudner2013}.

In finite systems, the topological index of a Floquet band can be calculated either using a discretized version of the integration in Eq.~\eqref{Ch} for the Chern number, or using the Bott index in Eq.~\eqref{bott-def}. As mentioned before, the Bott index calculation converges much more rapidly to the thermodynamic limit result with increasing system size \cite{Fukui2005}. How rapidly the critical value $A^*$ (obtained using the Bott index) for the topological transition converges depends on whether the momentum at which the gap closes in the thermodynamic limit is present in the discrete Brillouin zone of the finite system. In Fig.~\ref{fig-static}(b), we plot results for the critical value $eaA_L^{\ast}$ obtained for the first topological transition in systems with $L_1=L_2\equiv L$ as a function of $L$ (for $\Omega/J=7$). Only lattices in which $L=3\iota$ ($\iota\in \mathcal{\mathbb{Z}}$) contain the $K$ and $K'$ points (are {\tmem{commensurate}), where the Berry curvature is concentrated near the transition. They can be seen to produce a critical value that is system size independent starting from $L=3$. On the other hand, lattices with $L=3\iota+1$ and $L=3\iota+2$ exhibit a power-law approach of the critical value to the thermodynamic limit result [see Fig.~\ref{fig-static}(c)]. Whether $K'$ is included in the discrete Brillouin zone of the finite system plays a fundamental role in the unitary dynamics of the Bott index studied in what follows.

\section{Dynamics of the Bott index in finite systems}\label{sec:dynbott}

As mentioned before, the Chern number (Bott index) is a constant of motion in translationally invariant systems in the thermodynamic limit. However, there is nothing preventing the Bott index from changing during unitary time evolutions in finite systems, even if those systems are translationally invariant. One can consider two extreme cases of dynamics: (i) In a sudden quench of a Fermi sea, the finite system size limits the resolution of reciprocal space. As a result, a quenched Fermi sea can develop an increasingly complicated Berry curvature with time, such that the Bott index calculated in finite systems may be strongly dependent on time (for times larger than the linear system size divided by the maximal group velocity) and system size \cite{Caio2015, Caio2016, Sacramento2014}. (ii) In a system driven at a high frequency with a slowly increasing driving term, the Bott index can change if the system is able to evolve adiabatically in a Floquet picture \cite{Kitagawa2010,Breuer1989}. A finite time scale for adiabatic evolution can exist only if the $\mathbf{k}$ point at which the gap closes in the Floquet Hamiltonian (in the thermodynamic limit) is absent in the finite system.

A note is in order about the computation of the Chern number in finite systems out of equilibrium. When the magnitude of the driving term is increased slowly (in a system driven at high frequency), states away from the band gap of the Floquet Hamiltonian mainly evolve adiabatically. If the new band that is generated (in the Floquet picture) with increasing the strength of the driving term changes topology, then the Berry curvature of the original Fermi sea will accumulate about the gap-closing point(s) and vary rapidly about it (them), in order to observe the no-go theorem. The computation of the Chern number using Eq.~(\ref{Ch}) in finite systems then becomes numerically unstable. The Bott-index formula in Eq.~(\ref{bott-def}) should be the one used to study those systems out of equilibrium.

In our numerical calculations, we turn on the vector potential smoothly [its magnitude is increased linearly from zero, $A(t)\propto t$] to drive the ground state of the static Hamiltonian into the ground state of the Floquet one with $\tmop{Ch} = 1$, shown in Fig.~\ref{fig-static}(a). We consider only driving frequencies greater than the bandwidth, so that the Floquet bands are ordered unambiguously. We work in a regime in which $ea \dot{A}(t) \ll \Omega$. In this regime, one can think of the evolution of the time-dependent state as being dictated by a slowly changing Floquet Hamiltonian, so that traditional concepts such as adiabaticity can be applied \cite{Breuer1989}.

\begin{figure}[!t]
  \includegraphics[width=0.99\columnwidth]{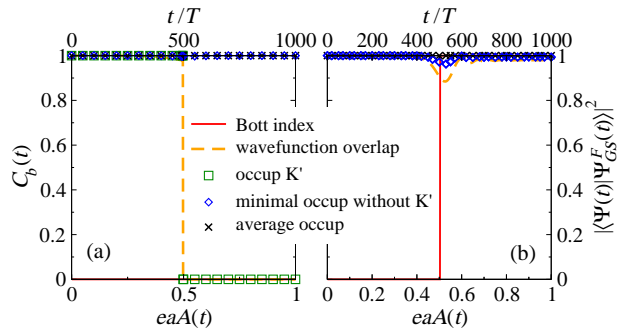}
  \caption{Bott index of the time-evolving wave function $|\Psi(t)\rangle$, and overlap between $|\Psi(t)\rangle$ and the ground state of the instantaneous Floquet Hamiltonian $|\Psi^F_{GS}(t)\rangle$ as a function of $eaA(t)$ (bottom labels) at stroboscopic times $t$ (top labels). We also show how the lower band of the instantaneous Floquet Hamiltonian is occupied in the time-evolving state. Specifically, we plot the occupation at the $K'$ point, the lowest occupation of any $\mathbf{k}$ state in the lower band excluding the $K'$ point, as well as the average occupation of the $\mathbf{k}$ states in the lower band. (a) Results for a $150 \times 150$ (commensurate) lattice. (b) Results for a $151 \times 151$ (incommensurate) lattice. Note that the latter does not contain the $K'$ point, so no result is reported for its occupation. In both lattices, the magnitude of the electric field is ramped up linearly from 0 to $eaA = 1$ in 1000 driving periods, for $\Omega = 7 J$ [corresponding to the black dot in Fig.~\ref{fig-static}(a)].}\label{fig-bott-occ}
\end{figure}

The Bott indexes of two time-evolving Fermi seas in which the magnitude of the electric field is slowly ramped from $eaA = 0$ to $1$ are plotted in Figs.~\ref{fig-bott-occ}(a) and~\ref{fig-bott-occ}(b) as a function of $eaA(t)$ (bottom labels) at stroboscopic times $t$ (top labels) for $\Omega = 7 J$. Figure~\ref{fig-bott-occ}(a) shows results for a $150 \times 150$ (commensurate) lattice, while Fig.~\ref{fig-bott-occ}(b) shows results for a $151 \times 151$ (incommensurate) lattice. The Bott index of the commensurate lattice [Fig.~\ref{fig-bott-occ}(a)] observes the no-go theorem, namely, it is conserved during the dynamics. On the other hand, the Bott index of the incommensurate lattice [Fig.~\ref{fig-bott-occ}(b)] changes from 0 to 1 when the magnitude of the electric field exceeds the critical value $A^*$ in Fig.~\ref{fig-static}(a). Namely, the initial topologically trivial Fermi sea evolves into a topologically nontrivial state under unitary dynamics. A first insight into the origin of the different behavior of the Bott index in those two lattices can be gained by studying the overlap between the time-evolving wave function and the instantaneous Floquet ground state, also shown in Fig.~\ref{fig-bott-occ}. For the commensurate lattice [Fig.~\ref{fig-bott-occ}(a)], that overlap is essentially 1 (near adiabatic evolution) up to about $A^*$, but then, when $A(t)$ becomes larger than $A^*$, the overlap vanishes, and the time-evolving state becomes orthogonal to the instantaneous (topologically nontrivial) Floquet ground state. On the other hand, for the incommensurate lattice [Fig.~\ref{fig-bott-occ}(b)], the overlap remains close to 1 (near adiabatic evolution) at all times. The smallest overlaps occur about $A^*$, but they are still higher than 0.8 and can be made arbitrarily close to 1 by decreasing the ramp speed.

For the commensurate lattice in Fig.~\ref{fig-bott-occ}(a), we also plot the occupation of the $K'$ point of the lower band of the instantaneous Floquet Hamiltonian in the time-evolving state, obtained by computing
 $\left|\langle u (K',t) | u_{GS}^F (K',t) \rangle \right|^2$, 
 where $|u (K',t)\rangle$ is the time-evolving wave function at time $t$ at $K'$ and $|u_{GS}^F(K',t)\rangle$ is the ground-state wave function of the instantaneous Floquet Hamiltonian at time $t$ at $K'$. The occupation of the $K'$ point can be seen to vanish when the magnitude of the vector potential exceeds $A^*$. This is the reason behind the vanishing of the overlap between the time-evolving wave function and the instantaneous Floquet ground state and, ultimately, behind the conservation of the Bott index. The next lowest occupied $\mathbf{k}$ state of the Floquet ground-state band, also shown in Fig.~\ref{fig-bott-occ}(a), is very close to 1. Namely, all but the $K'$ point evolve (nearly) adiabatically during the dynamics. As a result, the arithmetic mean of the occupation of $\mathbf{k}$ states of the Floquet ground-state band in the time-evolving state is very close to 1 [see Fig.~\ref{fig-bott-occ}(a)]. 

For the incommensurate lattice in Fig.~\ref{fig-bott-occ}(b), for which there is no vanishing gap in the Floquet Hamiltonian, the minimally occupied $\mathbf{k}$ state of the Floquet ground-state band during the dynamics is very close to 1 at all times, with the largest departure from 1 occurring when the magnitude of the vector potential is about $A^*$ (similarly to what is seen for the wave-function overlaps). As for the wave-function overlaps, the final occupation can be arbitrarily close to 1 if the ramp speed is decreased (the gap provides a well-defined time scale for adiabaticity). In the incommensurate lattice [Fig.~\ref{fig-bott-occ}(b)], the average occupation of $\mathbf{k}$ states of the Floquet ground-state band in the time-evolving state can also be seen to be very close to 1. We should stress that the magnitude of the vector potential at which the Bott index jumps in Fig.~\ref{fig-bott-occ}(b) and the overlap vanishes in Fig.~\ref{fig-bott-occ}(a) can change if one changes the ramping speeds. However, it converges to ${A}^{\ast}$ when $ea \dot{A} \ll \Omega$.

\begin{figure}[!t]
  \includegraphics[width=\columnwidth]{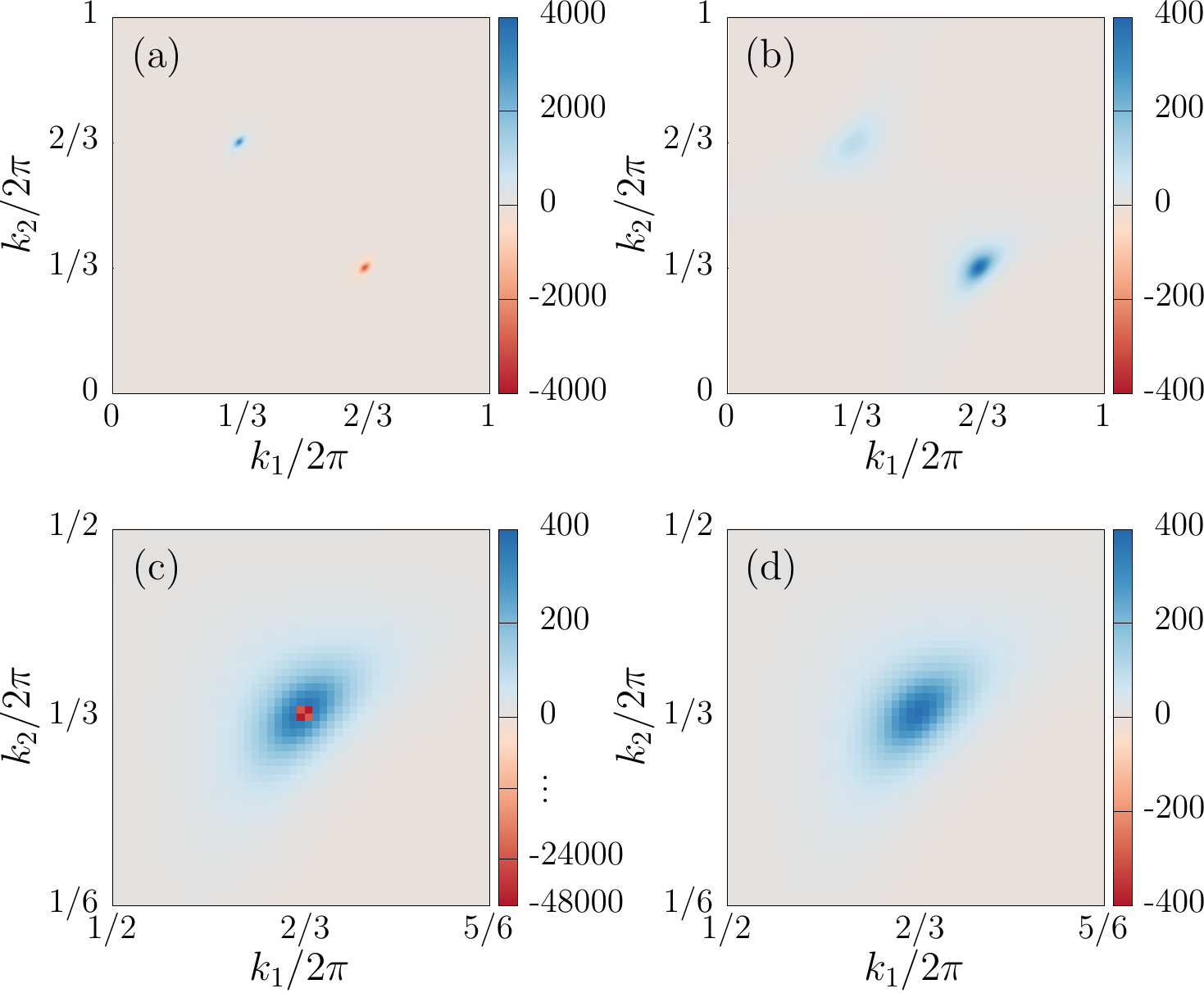}
  \caption{Berry curvature in the static, Floquet, and time-evolved Fermi seas of finite systems. The discrete $\mathbf{k}$ space is displayed in coordinates of the primitive reciprocal lattice vectors with unit lattice constants. $K$ and $K'$ are located at $(k_1, k_2) = \left( 2 \pi / 3, 4 \pi / 3 \right)$ and $\left( 4 \pi / 3, 2 \pi / 3 \right)$, respectively. (a) Berry curvature of the static, topologically trivial, ground state in a lattice with $L=900$. The signs are positive around $K$ and negative around $K'$. (b) Berry curvature of a Floquet, topologically nontrivial, ground state ($eaA = 1$ and $\Omega = 7 J$) in a lattice with $L=900$. All signs around $K$ and $K'$ are positive. (c) Berry curvature about $K'$ in a time-evolved commensurate system with $L=150$. Note the accumulation of negative Berry curvature at $K'$. The four data points in the center forming a $2 \times 2$ square are negative, opposite to all the surrounding points. (d) Berry curvature about $K'$ in a time-evolved incommensurate system with $L=151$.  All signs around $K'$ are positive. In (c) and (d), we show results from the time evolution of an initial topologically trivial Fermi sea after the magnitude of the electric field is ramped up linearly from $eaA = 0$ to 1 (with $\Omega = 7 J$) in 8000 driving periods. Here the ramp is 8 times slower than that in Fig.~\ref{fig-bott-occ} to make the Berry curvature about $K'$ indistinguishable (in the scale of these plots) between (b) and (d).}\label{fig-berry}
\end{figure}

Next, we study the Berry curvature of the static and Floquet ground states, as well as of the time-evolved Fermi seas in finite systems. We compute them from each term in Eq.~(\ref{bott-single-body}), dividing by the area element $\delta k^2$. The Berry curvature of the static, topologically trivial ground state is shown in Fig.~\ref{fig-berry}(a). In this case, the Berry curvature is mostly zero everywhere in the band and then large and positive (negative) about the $K$ ($K'$) point. This results in a vanishing Chern number. Figure~\ref{fig-berry}(b) shows the Berry curvature of a topologically nontrivial Floquet ground state (corresponding to $eaA = 1$ and $\Omega = 7 J$). In this case, the Berry curvature is once again mostly zero everywhere in the band, but then it is large and positive about the $K$ and $K'$ points (larger about $K'$). This results in $\tmop{Ch} = 1$.

As previously discussed, when one increases the magnitude of the electric field in our driven systems from zero, the first topological transition in the Floquet Hamiltonian occurs via a band-gap closing at $K'$, as a result of which the Chern number changes from 0 to 1. As hinted by our results in Fig.~\ref{fig-bott-occ}, something fundamentally different happens to the Berry curvature about that $K'$ point in commensurate and incommensurate lattices when one evolves unitarily the topologically trivial Fermi sea of the static Hamiltonian by slowly ramping up the magnitude of the electric field. This is shown in Figs.~\ref{fig-berry}(c) and~\ref{fig-berry}(d), respectively. In the time-evolved state of the commensurate lattice [Fig.~\ref{fig-berry}(c)], the Berry curvature close to (but not at) $K'$ is very similar to that in the Floquet Hamiltonian [Fig.~\ref{fig-berry}(b)]. However, at $K'$ the Berry curvature in the former is very large and negative, in contrast to the positive Berry curvature in the Floquet Hamiltonian. This is how the Bott index remains zero during the dynamics. On the other hand, in the time-evolved Fermi sea of the incommensurate lattice [Fig.~\ref{fig-berry}(d)], the Berry curvature about $K'$ is indistinguishable from that in the Floquet Hamiltonian [Fig.~\ref{fig-berry}(b)].

\begin{figure}
  \includegraphics[width=0.77\columnwidth]{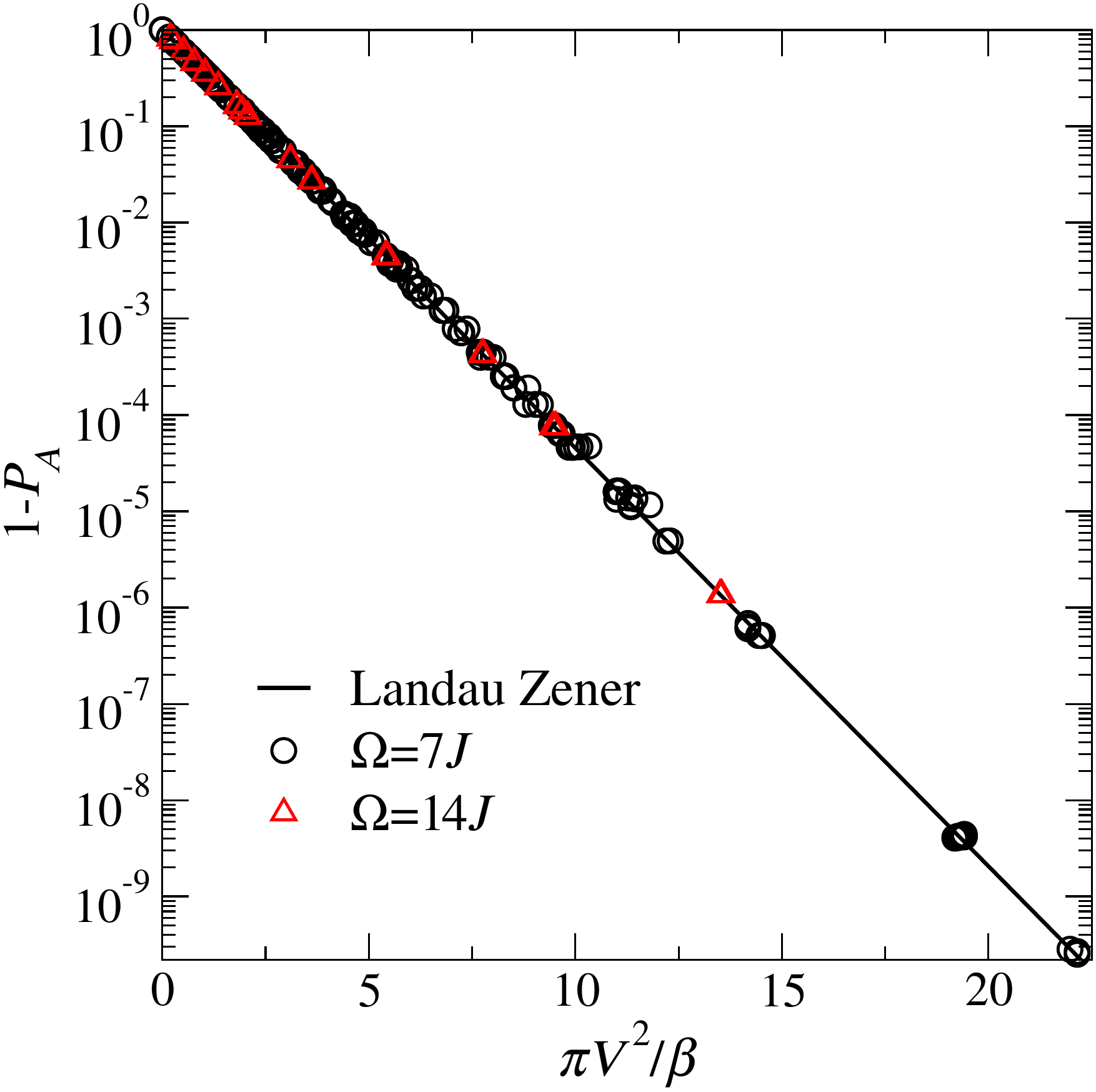}
  \caption{Collapse of the final occupation of momentum states near $K'$ to the Landau-Zener prediction (straight line). $\beta$ and $V$ are extracted from $\hat H_F [A(t)]$ before and after the gap closing. In all simulations, the magnitude of the electric field is turned on linearly from $eaA = 0$ to 1. We show results for several ramping rates (six), lattice sizes (ten, in which $L$ ranges from 600 to 16384), and values of $\mathbf{k}$ (about ten for each lattice size and ramping rate).}\label{fig-lz}
\end{figure}

As mentioned before, our study focuses on a regime in which the evolution of the time-evolving state is effectively dictated by a slowly changing Floquet Hamiltonian. Close to $K'$, the evolution is essentially a Landau-Zener problem \cite{Zener1932, Privitera2016, Hu2016,Breuer1989,Breuer1990}, in which the ground state evolves under a level-crossing Hamiltonian
\begin{equation}\label{eq:LZHam}
 \hat H_\mathbf{k} (t) \sim \beta_\mathbf{k} t \tilde{\sigma}_z + V_\mathbf{k}  \tilde{\sigma}_x .
\end{equation}
The Hamiltonian~\eqref{eq:LZHam} is written in the basis of the level-crossing eigenstates, which are the eigenstates of $\tilde{\sigma}_z$. $\beta_\mathbf{k}$ gives the rate at which the level-crossing point is passed, and $V_\mathbf{k}$ is the perturbing off-diagonal term, which is zero at $K'$ at the time at which the magnitude of the vector potential is $A^{\ast}$. 

The probability $P_{A}$ to remain adiabatic, i.e., in the ground state of the final Floquet Hamiltonian, is given by the Landau-Zener formula
\begin{equation}
  P_A (k) = 1 - \exp \left( - \frac{\mathpi | V_\mathbf{k} |^2} {\beta_\mathbf{k}} \right).
\end{equation}
The Landau-Zener parameters are extracted from $\hat{H}_F(A)$ calculated shortly before and shortly after $A$ becomes equal to $A^*$, $A_{\pm}$, which are $\Delta t$ away from each other (we used $eaA_{\pm} = eaA^{\ast} \pm 0.0015$ for the results shown). The Floquet Hamiltonians for those two values of $A$ allow us to extract both $\beta_{\mathbf{k}}$ from $[\hat{H}_{F,\mathbf{k}}(A_{+}) - \hat{H}_{F,\mathbf{k}}(A_{-})]/\Delta t$ and $V_{\mathbf{k}}$ from $[\hat{H}_{F,\mathbf{k}}(A_{+}) + \hat{H}_{F,\mathbf{k}}(A_{-})] / 2$. Results obtained for $1-P_A$, for several values of $\mathbf{k}$, lattice sizes, and ramping rates are plotted in Fig.~\ref{fig-lz} vs $\mathpi | V_\mathbf{k} |^2/\beta_\mathbf{k}$. They exhibit an excellent collapse to the Landau-Zener prediction.

The different behaviors between commensurate and incommensurate and, ultimately, finite- and infinite-size lattices have a geometric interpretation. An infinite-size lattice has a continuous Brillouin zone. The eigenstates of a filled band can be mapped onto a closed surface. The topological index of a given band reflects the topological charges enclosed by it \cite{Xiao2010,Hasan2010}. In a time-evolving traceless two-band model, one can map each state spinor $\left[ \cos \frac{\theta_\mathbf{k}(t)}{2}, \sin \tfrac{\theta_\mathbf{k}(t)}{2} \mathe^{\mathi \phi_\mathbf{k}(t)} \right]^{\small{T}}$ onto a point in three dimensions, with polar coordinates $[| E_\mathbf{k} (t) |, \theta_\mathbf{k} (t), \phi_\mathbf{k} (t)]$, where $E_\mathbf{k} (t)$ is the energy of the state in the instantaneous (Floquet in our case) Hamiltonian. The topological charge, corresponding to a band-touching point, sits at the origin. Under a slow evolution, the surface ``follows'' the time-dependent Hamiltonian $\hat{H}_\mathbf{k}(t)$; that is, it shifts and deforms. But whenever a patch of the surface moves close to the origin, the dynamics of that patch freezes due to vanishing $E_\mathbf{k}$. Consequently, topological charges always stay on the same side of the surface corresponding to the original energy band. Hence, the Chern number of such a filled band cannot change under unitary time evolutions, $\mathd \tmop{Ch} (n, t)/\mathd t = 0$. Intuitively, this geometric picture should generalize to higher dimensions for systems with more than two bands.

For a finite system, one has a discrete set of points rather than a closed surface. Therefore, the topological charge can easily ``leak'' through during the dynamics. This is allowed to happen unless the critical $\mathbf{k}$ point(s) at which $E_\mathbf{k} (t)$ vanishes is present to prevent it, such as $K'$ in our commensurate lattices. In general, the critical $\mathbf{k}$ points are model dependent; for example, they can depend on the presence of disorder and boundary conditions. For the transition to the $\tmop{Ch} = - 2$ phase in Fig.~\ref{fig-static}(a), the location of the three band touching points depends on both the strength of the staggered sublattice potential and the driving frequency. In such situations, one usually will not find that the Chern number is conserved during dynamics in finite lattices.

\section{Summary}\label{sec:summary}

Using elementary methods, we proved that the Bott index (generally used to study systems that lack translational symmetry) is equivalent to the Chern number in translationally invariant systems in the thermodynamic limit. As a byproduct of our proof, we showed that, when written in momentum space, the Bott index is nothing but the Chern number introduced in Ref.~\cite{Fukui2005} for finite translationally invariant systems. 

We used the Bott index in finite honeycomb lattices with periodic boundary conditions to determine the topological phase diagram of the Floquet Hamiltonian of driven spinless fermions with nearest-neighbor hoppings and a staggered potential. We then studied the dynamics of initial topologically trivial Fermi seas when the driving term is slowly ramped through a topological phase transition. We showed that while the Bott index is conserved in commensurate lattices, those that contain the $\mathbf{k}$ point at which the gap closes in the Floquet Hamiltonian, it is not conserved in incommensurate lattices. The latter behavior was the one observed in systems with open boundary conditions~\cite{DAlessio2015}. We argued that, in incommensurate lattices, adiabatic dynamics allows the Bott index to change at the critical value computed for the Floquet Hamiltonian. Hence, regardless of the no-go theorem for the thermodynamic limit \cite{DAlessio2015}, topological phase transitions can occur in finite translationally invariant systems provided there is a well-defined adiabatic limit for the effective Floquet dynamics.

\begin{acknowledgments}
We are grateful to Luca D'Alessio for many illuminating discussions at early stages of this project. Y.G. would like to thank Jiabin Yu, Rui-Xing Zhang and Qing-Ze Wang for helpful discussions. This work was supported by the Office of Naval Research, Grant No.\ N00014-14-1-0540. The computations were done at the Institute for CyberScience at Penn State.
\end{acknowledgments}

\bibliography{Bott-index-YG}

\end{document}